\documentstyle[12pt,twoside,fleqn,npa]{article}

\input{psfig.sty}

\newcommand{\AmS}{{\protect\the\textfont2
  A\kern-.1667em\lower.5ex\hbox{M}\kern-.125emS}}

\title{Response function beyond mean field of neutron-rich nuclei}

\author{G. Col\`o, P.F. Bortignon and R.A. Broglia\address{Dipartimento di 
        Fisica, Universit\`a degli Studi, \\ 
        and INFN, Via Celoria 16, 20133 Milano (Italy)}
       }

\begin{document}
\maketitle

\begin{abstract}
The damping of single-particle and collective motion in exotic isotopes is a
new topic and its study may shed light on basic problems of nuclear
dynamics. For instance, it is known that nuclear structure calculations 
are not able, as a rule, to account completely for the empirical 
single-particle damping. In this contribution, we present 
calculations of the single-particle self-energy in the case of the 
neutron-rich light 
nucleus $^{28}$O, by taking proper care of the continuum, and we show that
there are important differences with the case of nuclei 
along the valley of stability. 
\end{abstract}

\section{Introduction}\label{intro}

A useful approximation to describe atomic nuclei is in terms of an average 
potential in which neutrons and protons move independently of each other. 
This average potential is not a static one, and undergoes conspicuous 
fluctuations which give rise to collective oscillations of the system. 
The coupling of the particle motion to the collective modes is at the origin 
of the energy dependence of the average potential 
(cf., e.g., Refs.~\cite{Mah85,Mah91}). The empirical energy-dependent average 
potential is also able to account (at positive energy) for the behavior of a 
particle impinging on the nucleus in a scattering experiment, as it 
includes an imaginary part which corresponds to the absorptive scattering channels.
Within the so-called nuclear structure approach, the
Hartree-Fock (HF) mean field can thus be viewed as a starting point to which 
corrections can be added, the leading ones being associated with processes 
in which a nucleon excites 1 particle-1 hole (1p-1h)  
pairs -- preferably correlated pairs, that is, nuclear collective vibrations.
Such corrections have been extensively studied for nuclei along the valley
of stability. As a rule one finds
that this approach is able to account only for about 50\% of the 
empirical damping of single-particle motion~\cite{Ber83}. 
This is still an unresolved issue and points to the limits of accuracy of our
understanding of the nuclear dynamics. 

In keeping with these facts, it is quite natural to ask the question whether
in the recently discovered exotic nuclei (for instance, in systems with
larger neutron excess compared to the stable isotopes) the situation
described above is changed or not. This is one motivation for the present
work. An even stronger motivation comes from the fact that, as it is well 
known~\cite{Ber83}, the coupling of the single-particle motion to nuclear vibrations 
is responsible for most
of the spreading widths of giant resonances in stable nuclei. In the case
of exotic nuclei, the importance of the coupling with the continuum, that
is, of the escape width of both
the single-particle states and collective states has been stressed in many
works. On the other hand, not much work has been done to calculate the 
spreading width due to 
coupling with more complicated configurations (2p-1h for single-particle
states, 2p-2h for giant resonances). 

In a previous work~\cite{Ghi96}, we have calculated the quadrupole response
in the neutron-rich nucleus $^{28}$O by taking into account both the escape
and spreading width. We have reached the conclusion that the spreading width
is as important as the escape width, at least in the example 
considered. Some of the results of Ref.~\cite{Ghi96} were considered to be
model-dependent, so that a more refined calculation is highly desirable. 
In fact, so far one has been able to 
take properly into account the continuum part of the
single-particle spectrum in calculations at the 1p-1h level and ``more
refined calculation'' means therefore, within the present 
context, a proper treatment of the continuum also at the 2p-2h level. With this
purpose we performed calculations in which the self-energy of particles
(holes) is obtained by coupling with 2p-1h
(2h-1p) configurations and for the first time in the study of exotic
nuclei, we properly treat the continuum at this 2p-1h (2h-1p) level. 

Preliminary results are reported in this
contribution. In a previous work~\cite{Col98}, considerations
based on analytical expressions had been presented and we confirm here the
soundness of these considerations by means of some numerical findings.  

\section{Imaginary part of the single-particle self-energy in exotic nuclei}
\label{imag}

The imaginary part of the single-particle self-energy felt by a nucleon  
in the presence of an $A$-particle bound system has been one of the main 
concerns of many studies in nuclear structure, both experimental and
theoretical. When the nucleon is at positive energy and its wave function 
has ingoing wave boundary conditions, this imaginary part corresponds to 
the imaginary part of the optical potential. Whereas for heavy ions the
dominant contribution to the optical potential comes from transfer phenomena
and this can be accurately calculated~\cite{Bro91}, in the case of nucleon 
scattering the theoretical understanding 
of the optical potential is still a rather open problem. In fact, there is 
not a clear-cut explanation why, for instance, the optical potential
calculated 
by starting from nuclear matter studies and using a local density 
approximation to deal with finite nuclei, reproduces more or less the 
strength of the empirical optical potential; and why, on the other hand, 
within the above mentioned nuclear structure approach, only about one half 
of the potential strength can be accounted for.

\begin{figure}
\psfig{figure=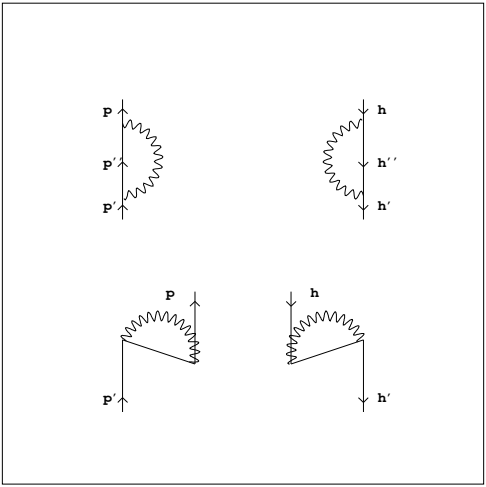,height=8.0cm,width=8.0cm}
\caption{The diagrams corresponding to the leading terms of the 
particle (left column) or hole (right column) self-energy. 
\label{fig:2diagrams}}
\end{figure}

Within this latter approach~\cite{NVM79,Ber79}, the optical potential is calculated 
by adding to the HF mean field (V$_{HF}$) the contribution $\Delta$V coming 
from the excitation of 2p-1h states (mainly correlated, that is, 1p-1 
collective vibration). Only $\Delta$V gives rise to an imaginary part in 
the single-particle potential. In particular, only the upper two
diagrams associated with $\Delta$V and shown in Fig. 1, are relevant. 
We report the analytic expression for the upper left 
diagram, contributing to the particle self-energy, which reads
\begin{eqnarray}
 Im \Delta {\rm V}_{lj}(r,r^\prime ; \omega) & = & \sum_{l^\prime,j^\prime,
 \lambda} \int d\omega_\lambda\ (-\pi)v(r)v(r^\prime) 
 {\tilde u_{l^\prime j^\prime}^{(\omega-\omega_\lambda)}(r)\over r}
 {\tilde u_{l^\prime j^\prime}^{(\omega-\omega_\lambda)}(r^\prime)
 \over r^\prime} \ \times \hfill\nonumber \\
 \ & \times & \delta\varrho^{(\omega_\lambda)}(r)
 \delta\varrho^{(\omega_\lambda)}(r^\prime) 
 { \langle lj \vert\vert Y_\lambda \vert\vert l^\prime j^\prime \rangle^2
 \over 2j+1 }. 
\label{ImV}\end{eqnarray}
The quantum numbers $l,j$ ($l^\prime,j^\prime$) refer to the initial 
(intermediate) state, ${\tilde u}(r)$ are radial wave functions of the 
particles, $\delta\varrho(r)$ are transition densities of 1p-1h pairs 
coupled to multipolarity $\lambda$ and $v(r)$ is 
the p-h interaction derived from the effective force from which the HF mean 
field is determined 
(in our case, we have used the Skyrme force SIII). The link between 
Eq.~(\ref{ImV}) and the empirical imaginary part of the optical potential 
$W$ is as follows, 
\begin{equation}
 W(r,r^\prime ; \omega) = \sum_{lj} {2j+1\over 4\pi} Im \Delta
 V_{lj}(r,r^\prime ; \omega).
\label{opt_tot_nl}\end{equation}
The real part is obtained in a similar way, or through 
dispersion relation techniques (see Ref.~\cite{Mah85,Mah91} and references 
therein). A procedure exists to connect the non-local potential with 
its local equivalent, which is the quantity to be compared with 
the empirical parametrizations or to be given as an input in 
a reaction calculation, e.g, 
of DWBA type. 

The novel feature of Eq.~(\ref{ImV}), in comparison with what has been 
used in the past, is that we treat properly the continuum also at the 
level of the 2p-1h doorway states. In the calculations performed in the 
seventies in well-bound nuclei like $^{208}$Pb, discrete particle states were
employed as the whole system was set in a box and an averaging parameter
was employed in Eq.~(\ref{ImV}) to ensure the match of the initial 1p and
intermediate 2p-1h energies (see Ref.~\cite{Mah85} and references therein). 
This procedure is satisfactory for well-bound systems as we 
have been able to confirm by making use of Eq.~(\ref{ImV}) for a test 
calculation in $^{208}$Pb where we have reproduced essentially the 
results of Ref.~\cite{Ber79}. For systems with loosely bound nucleons, 
where the correct treatment of the continuum is more important, 
the use of discrete particle states is not able to reproduce 
the results that we 
are going to illustrate in Sec.~\ref{res}.  

Before that, we use Eq.~(\ref{ImV}) to get 
some insight in the qualitative low-energy behavior of the imaginary part 
of $\Delta$V (by ``low-energy'' we mean for values of the energy 
$\omega$ close to the particle emission threshold $\omega_{th}$). 
We use the fact that in the case of 
nuclei with loosely bound neutrons, the low-energy part of the (1p-1h) 
multipole response is characterized by a pronounced ``threshold effect'', 
that is, by a sudden increase of the strength function $S(\omega)$ above 
$\omega_{th}$~\cite{Hiro,Col98}. On ground of simple arguments, it is 
expected that
\begin{equation}
 S(\omega) \sim (\omega-\omega_{th})^{l^\prime+1/2},
\label{sthr}
\end{equation}
where $l^\prime$ is the orbital angular momentum of the particle states 
contributing to $S(\omega)$. 

Let us consider a single term of the sum appearing in eq.~(\ref{ImV}) and 
fix $r=r^\prime$. A plane-wave approximation for $\tilde 
u_{l^\prime j^\prime}^{(\omega-\omega_\lambda)}(r)$ suggests that it 
contributes with a factor $k_{part}^{l^\prime + 1/2}$ if 
$k_{part}=\hbar^{-1}\sqrt{2m(\omega-\omega_\lambda)}$ is close to zero.
Adding the condition on the normalization 
of the radial transition densities given by
\begin{displaymath}
 |\int dr\ r^{2+\lambda}\delta\varrho^{(\omega_\lambda)}(r)|^2 = 
 S(\omega), 
\end{displaymath}
with the asymptotic behaviour of the strength recalled in Eq.~(\ref{sthr}), 
we obtain 
\begin{equation}
 Im \Delta V \sim \int_{\omega_{th}}^{\omega} d\omega_\lambda\ 
 (\omega-\omega_\lambda)^{l'+1/2} (\omega_\lambda-\omega_{th})^{l''+1/2}.
\label{approxDeltaV}\end{equation}
If $\omega$ is close to $\omega_{th}$, i.e., $\omega=
\omega_{th}+\delta$ and $\omega_\lambda=\omega_{th}+\delta/2$, 
we find that the imaginary part of the optical potential behaves 
approximately like $\delta^{l'+l''+2}$, that is, a fast increase just 
after threshold which has no counterpart in stable nuclei (where $Im \Delta
V \sim (\omega-\varepsilon_F)^n$, with $1\leq n\leq 2$~\cite{Mah91}) and 
has consequences on the {\em qualitative} 
features of scattering experiments 
with exotic beams. A realistically calculated $Im \Delta
V$ follows indeed this asymptotic behaviour and this is shown in the next
Section. 

\section{Results}\label{res}

We have chosen as an example of neutron-rich nucleus, the isotope 
$^{28}$O. It is double magic and the neutron separation energy (about 1 
MeV according to the HF calculation already mentioned, employing the SIII 
interaction) lies between the stable nuclei and the extreme cases of 
nuclei with a very
extended halo like $^{11}$Li. We must recall that  
a number of experimental evidences have been recently accumulated which 
point to the non-existence of this isotope as a bound 
system~\cite{Fau_o28}. Since 
essentially all calculations predict $^{28}$O to be bound, a strong point 
should be made about the necessity of improving our tools for nuclear 
structure studies, to face the outcome of the considerable experimental 
efforts made in the field of drip-line nuclei.

\begin{figure}
\psfig{figure=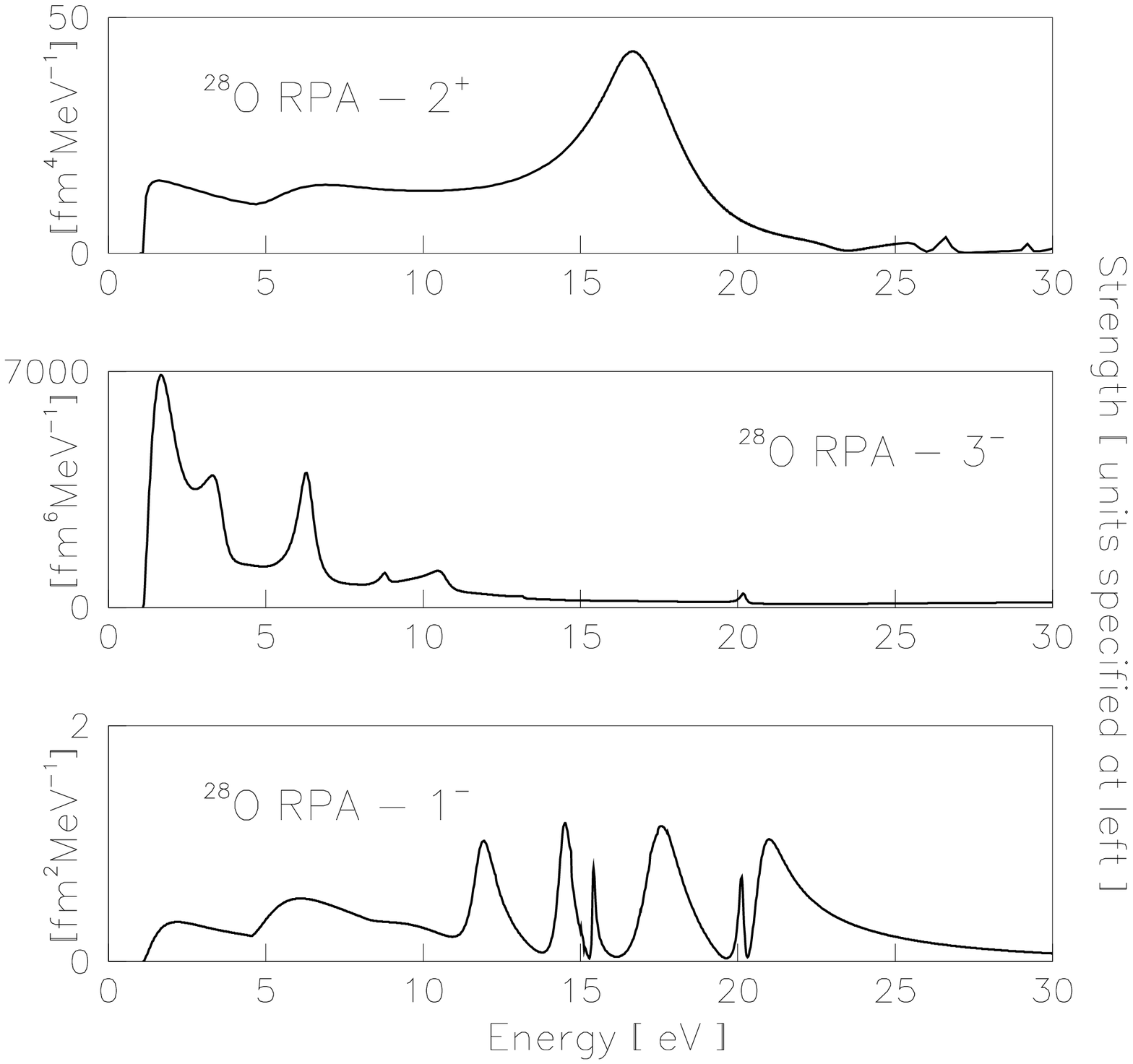,height=8.0cm,width=8.0cm}
\caption{Strength function for the isoscalar 2$^+$, 3$^-$ and isovector 1$^-$, 
calculated within continuum-RPA. \label{strengths}}
\end{figure}

We have calculated, in the case of neutrons, 
the quantity $W$ defined by 
Eq.~(\ref{opt_tot_nl}). First, we have solved the HF equations with the 
Skyrme interaction SIII on a radial mesh of 250 points with a 0.1 fm step. 
As a result, the levels shown in Fig. 1 of Ref.~\cite{Ghi96} are obtained. 
In particular, the highest occupied neutron orbital d$_{3/2}$ has a small 
binding energy of about 1.1 MeV while the protons are all bound by more than 
30 MeV. This large asymmetry between the neutron and proton mean fields is 
clearly due to the large neutron excess. Its consequences on the multipole 
response are quite dramatic as remarked first in Ref.~\cite{Hiro}. In fact, we have 
calculated~\cite{Ghi96,Col98} the strength function associated with 
isoscalar quadrupole, octupole and dipole operators, within the framework of 
continuum RPA in coordinate space~\cite{Liu76}. This theory is able to 
treat properly the particle states in the continuum, because the one-body 
propagator has an exactly computable form if expressed as function of 
$\vec r$, ${\vec r}^{\ \prime}$. We have kept 25 fm as the upper limit of the 
radial integrals while the radial mesh has a 0.5 fm step. The strength 
functions mentioned above are depicted in Fig.~\ref{strengths}. Their main 
feature is that they reflect how ``excess neutrons'' and ``core particles'' 
degrees of freedom are decoupled in systems like the one at hand. The 
low-energy part, although characterized by a strong enhancement just above 
the particle threshold, is not associated to any collective effect. 
What is enhanced so much to form a well-defined bump, are the
single-particle transitions from the d$_{3/2}$ neutron state to s-, p-, or 
d-states in the continuum: 
because of the small d$_{3/2}$ binding energy, the wave function of these
neutrons is so extended that its overlap with continuum wave functions is 
large and this makes the matrix elements of the multipole operators quite 
large. On the other hand, the higher energy region of the strength 
functions is characterized by states to which more ph configurations 
partecipate (neutron excess states play a predominant role but neutron core 
states are not negligible). Transitions from proton states are completely 
decoupled and lie above 30 MeV. 

From the continuum RPA calculation we can extract the radial transition 
densities $\delta\varrho^{(\omega_\lambda)}(r)$ which appear in 
Eq.~(\ref{ImV}). For each multipolarity $\lambda$ = 2$^+$, 3$^-$, 1$^-$, 
we take energy bins of 1 MeV for $\omega_\lambda$ and we associate to each
of them the appropriate $\delta\varrho(r)$ so that the integral $\int d\omega_\lambda$ 
can be performed. The continuum wave functions $\tilde u(r)$ are obtained 
by integrating the radial Schr\"odinger equation including the kinetic 
energy plus the HF potential at positive energy $\omega-\omega_\lambda$. 

\begin{figure}
\psfig{figure=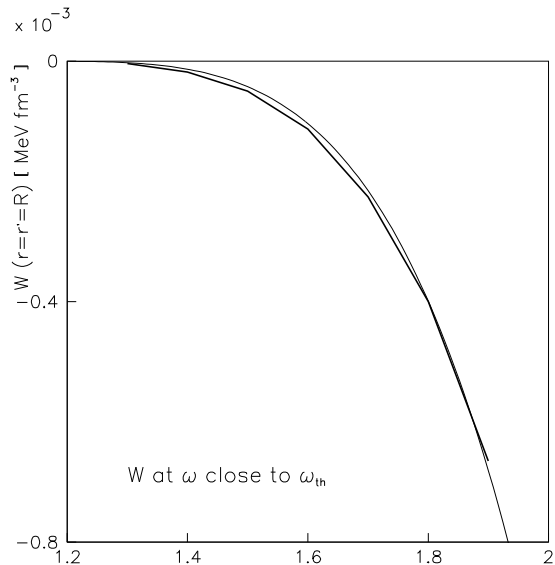,height=8.0cm,width=8.0cm}
\caption{Optical potential $W$ as defined in Eq.~(\ref{opt_tot_nl}), 
plotted in the case of $r=r'=3.2$ fm, as a function of the energy 
$\omega$ close to $\omega_{th}$ (thick line). In this low-energy region, 
the approximation $W\sim(\omega-\omega_{th})^4$ is valid (thin line).
\label{analytic}}
\end{figure}
  
We first study the low-energy asymptotic behavior of $W$ 
(Eq.~(\ref{opt_tot_nl})), to see whether it meets the 
expectations from Eq.~(\ref{approxDeltaV}). In Fig.~\ref{analytic}, we show 
(by means of thick line) the values of $W$ for $r=r'$ fixed approximately at 
the nuclear surface\footnote{Fig. 1 of Ref.~\cite{Col98} shows that the 
presence of nucleons with binding energies of a few MeV give rise to a very 
diffuse surface. We have taken here a value of 3.2 fm, where the density is 
approximately half of its central value.}. The thin lines is a function of the type 
$C(\omega-\omega_{th})^4$, with $C$ constant and chosen to get the best fit 
to the microscopic calculation. The good agreement between the two lines 
confirms the soundness of the argument of the last Section, about the
asymptotic behavior of the optical potential. 

\begin{figure}
\psfig{figure=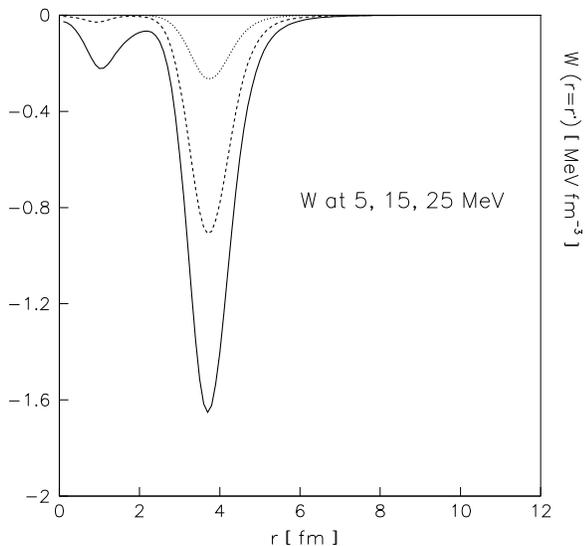,height=8.0cm,width=8.0cm}
\caption{Optical potential $W$ as defined in Eq.~(\ref{opt_tot_nl}), 
plotted in the case of $r=r'$, as a function of the radial distance for
three different values of the energy $\omega$ (dotted line: 5 MeV; dashed
line: 15 MeV; full line: 25 MeV). \label{opt_three_en}}
\end{figure}

In Fig.~\ref{opt_three_en}, $W(r,r'=r;\omega)$ is plotted as a function of 
$r$ for three different values of the energy $\omega$. As in stable nuclei, 
this function is surface peaked. On the other hand, the width of the peak 
is rather large, as can be expected if the nucleus has extended radial 
wave functions. All these characteristics of the imaginary part of the 
self-energy in exotic nuclei will be studied in more details in a 
forthcoming work. We can also notice that for $\omega$ = 25 
MeV a second bump, at the interior of the nucleus, shows up. This can be
interpreted as follows. In the calculation of the terms of the 
sum appearing in Eq.~(\ref{opt_tot_nl}), one performs integrals of the type 
$\int_{\omega_{th}}^\omega d\omega_\lambda$. If $\omega$ is large, 
one includes, through the transition densities, contributions from protons 
and these are of course peaked elsewhere
than on the nuclear surface, since the protons are concentrated in a much 
smaller region. In this respect, this interior peak is another example  
of the general statement according to which core particles and excess 
particles are decoupled in this light, exotic isotopes.

\section{Real part of the self-energy}\label{real}

The off-shell diagrams which are shown in the lower part of
Fig.~\ref{fig:2diagrams} contribute to the real part of the self-energy, in
addition to the two which appear above. It has been shown in the past (see
Ref.~\cite{Mah85} and references therein) that dispersion relation
techniques can be used to determine the values of Re $\Delta$V, once Im
$\Delta$V has been calculated. We do not recall here but the basic idea, by
using again the particle self-energy (diagrams in the left column of
Fig.~\ref{fig:2diagrams}). 

The imaginary part of the self-energy (Eq.~(\ref{ImV})) can be obtained as a
limit,
\begin{equation}
 Im \Delta {\rm V}_{lj}(r,r^\prime ; \omega) = Im \lim_{\eta \rightarrow 0}  
 \sum_{l^\prime,j^\prime,\lambda} \int d\omega_\lambda\ \int d\omega^\prime\ 
 {f \over \omega - (\omega^\prime + \omega_\lambda) + i\eta},
\label{limit}\end{equation}
where $f\equiv f_{lj;l^\prime,j^\prime,\lambda}(r,r^\prime ;
\omega,\omega^\prime, \omega_\lambda)$ denotes the product of the two
particle-vibration vertices. The integral over $d\omega_\lambda$ runs
from $\omega_{th}$ to $+\infty$. The integral over $d\omega^\prime$
can be extended from $-\infty$ to $+\infty$ and then the
contribution from the 2 particles-1 hole-1 phonon 
intermediate states (lower left diagram of
Fig.~\ref{fig:2diagrams}) is also included. 
This is just a formal extension if one takes the
imaginary part of the limit $\eta \rightarrow 0$, since this results in a 
delta-function $\delta(\omega^\prime-(\omega-\omega_\lambda))$ which rules
out the 2 particles-1 hole-1 phonon states contribution if $\omega$ is positive. But if
we consider the real part in the limit $\eta \rightarrow 0$, we easily get 
the expression for the real part of the self-energy, 
\begin{eqnarray}
 Re \Delta {\rm V}_{lj}(r,r^\prime ; \omega) & = & Re \lim_{\eta \rightarrow 0}  
 \sum_{l^\prime,j^\prime,\lambda} \int d\omega_\lambda\ \int d\omega^\prime\ 
 {f \over \omega - (\omega^\prime + \omega_\lambda) + i\eta} \hfill\nonumber
 \\
 \ & = & -{1\over\pi} P \int_{-\infty}^{+\infty} d\omega^\prime\ 
 {Im \Delta {\rm V}_{lj}(r,r^\prime ; \omega^\prime) \over \omega -
 \omega^\prime}.
\label{ReV}\end{eqnarray}

In the case of $^{28}$O, by using $Im \Delta {\rm V}_{lj}(r,r^\prime ;
\omega^\prime)$ calculated as it is described in Sec. 3, we have obtained
results also for the real part of the self-energy given by Eq.~(\ref{ReV}).
The problem we have addressed is actually the energy shift due to the
coupling with the states labelled by $\lambda,\omega_\lambda$ (which are
collective vibrations if $\omega_\lambda$ is around 15 MeV and are
single-particle transitions at lower energy). This shift is given exactly
by 
\begin{equation}
 \Delta E_{nlj} = \int dr d{r^\prime}\ r u_{nlj}(r)\ 
 Re \Delta {\rm V}_{lj}(r,r^\prime ; \omega)\ r^\prime u_{nlj}(r^\prime),
\label{shift}\end{equation}
where $\omega$ is usually fixed as the unperturbed (i.e., HF) energy 
$E_{nlj}^{(0)}$ of the level under consideration. The expressions for 
$Re \Delta {\rm V}_{lj}$ in the case of hole states are analogous to those
derived above for the case of particles. In the case of 
$^{28}$O calculated within the SIII-HF procedure, no unoccupied particle
states can be found at negative energy and the only particle resonance at 
positive energy is the f$_{7/2}$, so that the particle
spectrum consists essentially only of a smooth continuum. Therefore, we have
calculated the shift $\Delta E_{nlj}$ for neutron hole states, in particular
for the loosely bound d$_{3/2}$ orbital. 

We find a positive shift of 330 keV, to be compared with
$E^{(0)}$ = -1.1 MeV. This makes the nucleus more close to being
unbound and of course would in turn affect the strength function of the
multipoles we have considered in Sec. 3. As a conclusion to this part, 
the following two statements about the role of the real part of the
self-energy in neutron-rich nuclei can be made. (a) The particle shift of 
330 keV obtained for the loosely bound d$_{3/2}$ orbital forces us 
to ask questions about the reliability of simple mean field methods to
predict the single-particle energies and consequently the position of drip
lines, as well as the properties of the excited states. 
This is not the case for stable nuclei, since one usually obtains 
corrections of the order of less than 1 MeV to the energy of states which 
are usually 7-10 MeV bound. These corrections are 
important to give the correct density of states around the Fermi
energy and effective mass but do not alter the predictive power of theories
like the standard RPA in which these corrections are not taken into account. 
(b) The fact that the net result of 330 keV 
comes from a strong cancellation between the
two diagrams of the right column of Fig.~\ref{fig:2diagrams} (930 keV - 
600 keV), one has to wonder about the use of second RPA (as we did 
in~\cite{Ghi96}) where only the first of the two diagrammatic contributions 
is included. In conclusion, we believe that further research along this line
can address some basic problems of the present physics of nuclei with
large neutron excess.

\end{document}